\let\oldvec\vec% Store \vec in \oldvec (fix amsmath warning about vec)
\documentclass{llncs} % No specific formatting requirements by Elsevier.
\let\vec\oldvec% Restore \vec from \oldvec (fix amsmath warning about vec)

\usepackage[utf8]{inputenc}
\usepackage[T1]{fontenc}
\usepackage{url}
\usepackage{todonotes}

\usepackage[caption=false]{subfig}
\usepackage{amsmath,hyperref}
\usepackage[binary-units]{siunitx}

\usepackage[markup=underlined]{changes}

\begin{document}

\title{The Status of Quantum-Based Long-Term Secure Communication over the Internet}
%\title{The Status of Long-Term Secure Communication on the Internet via Quantum Cryptography}
%\title{Long-Term Secure Communication on the Internet via Quantum Cryptography}

\author{
Matthias Geihs\inst{1} \and
Oleg Nikiforov\inst{2} \and
Denise Demirel\inst{1} \and
Alexander Sauer\inst{2} \and
Denis Butin\inst{1} \and
Felix G\"unther\inst{3} \and
Gernot Alber\inst{2} \and
Thomas Walther\inst{2} \and
Johannes Buchmann\inst{1} 
}
\institute{
Kryptographie und Computeralgebra, TU Darmstadt, Germany\\
%\email{mgeihs@cdc.informatik.tu-darmstadt.de}
%\email{ddemirel@cdc.informatik.tu-darmstadt.de}
%\email{dbutin@cdc.informatik.tu-darmstadt.de}
%\email{buchmann@cdc.informatik.tu-darmstadt.de}
\email{\{mgeihs,ddemirel,dbutin,buchmann\}@cdc.informatik.tu-darmstadt.de}
 \and
Institut f\"ur Angewandte Physik, TU Darmstadt, Germany\\
%\email{oleg.nikiforov@physik.tu-darmstadt.de}
%\email{alexander.sauer@physik.tu-darmstadt.de}
%\email{thomas.walther@physik.tu-darmstadt.de}\\
%\email{gernot.alber@physik.tu-darmstadt.de}
\email{\{oleg.nikiforov,alexander.sauer,gernot.alber,thomas.walther\}\\@physik.tu-darmstadt.de}
 \and
Kryptographie und Komplexit\"atstheorie, TU Darmstadt, Germany\\
\email{guenther@cs.tu-darmstadt.de}
}
\maketitle

\begin{abstract}
Sensitive digital data, such as health information or governmental archives, are
often stored for decades or centuries. The processing of such data calls for long-term security.
Secure channels on the Internet require robust key establishment methods.
Currently used key distribution protocols are either vulnerable to future attacks based
on Shor's algorithm, or vulnerable in principle due to their reliance on computational
problems. Quantum-based key distribution protocols are information-theoretically secure
and offer long-term security. However, significant obstacles to their real-world use
remain. This paper, which results from a multidisciplinary project involving computer
scientists and physicists, systematizes knowledge about obstacles to and strategies
for the realization of long-term secure Internet communication from quantum-based
key distribution. We discuss performance and security particulars, consider the specific
challenges arising from multi-user network settings, and identify key challenges for
actual deployment.
\end{abstract}

%\todo[inline]{- Present version to Buchmann (14.9.)\\- Submission to \emph{Elsevier Journal of Computers \& Security}}
%\todo[inline]{All: Ganz zum Schluss \"uberpr\"ufen, ob alle Begriffe klar sind.}
%\todo[inline]{Word Count goal: between 5k and 10k. As of September 12th: 7,2k words.}
%\todo[inline]{kritik an QKD aufgreifen. hardware side channels --> gegenargument device independent QKD. keine richtige 'crypto'}
%\todo[inline]{no solution for it channel in classical crypto. carrying around hard disks with keys is not an alternative.}
%\todo[inline]{scenario, alice schickt etwas an bob, secure channel, zusammenhang QKD und OTP}
%\todo[inline]{existierende netzwerke erwaehnen.}
%\todo[inline]{kosten effizienz, existierende kabel benutzen?}
%\todo[inline]{alternatives: bounded storage, noisy channel}

%!TEX root = main.tex
\section{Introduction}

%Vast amounts of information are exchanged online every day.
The basis for exchanging a vast amount of information in today's world is the Internet --- a network that spans the earth and allows any two parties to communicate with each other.
If sensitive information is about to be transmitted (e.g., medical records or governmental documents), secure  connections must be established to protect confidentiality, integrity, and authenticity.
The most common method for establishing such a secure connection is to use the Transport Layer Security (TLS) protocol~\cite{TLS12}, which combines a key distribution protocol with a channel protocol.
%This protocol combines a key distribution protocol with a channel protocol in order to establish a secure channel between two communication partners.
First, the key distribution protocol is run to establish a common secret key unknown to a potential eavesdropper tapping the communication.
Then, the obtained key is used in the channel protocol to encrypt and authenticate the transmitted data, protecting its confidentiality and integrity.

Currently, the most commonly used key distribution protocol is based on the Diffie--Hellman key exchange~\cite{DHKE}.
Diffie--Hellman key exchange provides so called \emph{computational security}.
Its security is based on the assumption that computing discrete logarithms in certain groups is computationally infeasible (i.e., it would take a prohibitively long time for the computation to finish).
However, it has been shown that quantum computers can efficiently compute such discrete logarithms~\cite{shor} and, therefore, Diffie--Hellman key exchange is rendered insecure once quantum computers are available.
Recently, alternative key distribution protocols based on lattice cryptography have been proposed (e.g.,~\cite{newhope,frodo}) that are conjectured secure against quantum computers.
However, in principle, these protocols are potentially still vulnerable to computational attacks as their security is based on a computational problem.
Such computationally secure key distribution protocols achieve security only for a limited period of time, i.e., as long as attackers do not have sufficient computational power to break the security and obtain the exchanged secret keys.
%When this happens, the confidentiality of all transmitted data is lost.

An alternative to computational security is \emph{information-theoretic security}.
Information-theoretically secure components are not vulnerable to computational attacks (e.g., brute force attacks) and therefore provide long-term security against them.
Channels providing long-term confidentiality require information-theoretically secure key distribution and encryption.
%Such an encryption requires the secret key to be of the same length as the transmitted data.
%The One-Time Pad (OTP) encryption scheme \cite{shannon1949} is optimal in that sense.
The integrity demands for such a channel are usually only temporary (computational), that is, it is sufficient to guarantee integrity while the data is in transit.
While substantial effort has been made to define, understand, and construct computationally secure channels (originating, e.g., from~\cite{TISSEC:BelKohNam04}), a thorough understanding of how to construct information-theoretically secure channels achieving standard security goals of confidentiality and integrity as well as replay and reordering protection is still lacking.
For information-theoretic encryption, one-time pad (OTP) encryption~\cite{shannon1949} is an optimal solution.
We remark, however, that any information-theoretically secure encryption scheme requires the secret key to be of the same length as the encrypted data.
There exist several candidates for information-theoretically secure key distribution.
A naive approach is to distribute keys using a trusted courier that physically delivers a generated key stored on a hard drive.
This approach, however, suffers from obvious disadvantages with regards to practicability because it requires hours of traveling by the courier, which is far too long for most practical applications.
Other approaches for information-theoretically secure key distribution are protocols in the bounded storage model~\cite{bsm} or the noisy channel model~\cite{nsm}.
However, it is currently unclear how these models can be realized in practice~\cite{ltconfsurvey}.
Currently, the most promising approach for information-theoretically secure key distribution is Quantum Key Distribution (QKD). The security of QKD is based on the laws of quantum physics and its feasibility has already been demonstrated in many field tests~\cite{Chapuran2009,Elliott2005,Peev2009,Sasaki2011}.
However, there are still several challenges in order to realize QKD-based long-term secure communication on the Internet.
The performance and security of implementations of QKD protocols is still an issue.
Furthermore, techniques need to be developed that allow using QKD-based secure channels in distributed networks with many users.

In this work, we first discuss the current state of quantum-based key distribution protocols with regards to performance and security (Sec.~\ref{sec.quantumke}).
We then discuss the current state of enabling QKD in multi-user networks (Sec.~\ref{sec.qkehubs}).
Finally, we summarize the challenges yet to be addressed in order to realize QKD-based long-term secure communication on the Internet (Sec.~\ref{sec.outlook}).

%!TEX root = main.tex

\section{Quantum-based Key Distribution}
\label{sec.quantumke}

%There are different approaches to build a quantum computer. However, all of the existing prototypes have only low computational capacity or they are designed for specialized applications (D-Wave can solve only optimization problems)\cite{}.

We now describe the state of the art for QKD.
We first explain relevant concepts of quantum physics, then we categorize and summarize prominent QKD protocols. Next, we compare the performance of the protocols, and finally we discuss security models and attacks on protocol implementations.

%Describe quantum-based network infrastructure.
%Security based on laws of quantum physics.
%Requires channels, hubs, repeaters, trusted nodes.

\subsection{Quantum Physics Background}

%QKD protocols are based on laws of quantum physics and involve preparing and transmitting quantum states.

%\subsubsection{Laws of Quantum Physics.}
QKD protocols rely on fundamental laws of quantum physics: the typical change of state of a quantum object after a measurement and the impossibility to copy a quantum state without disturbing the state of the original particle.
Security of QKD protocols relies on the fact that a potential eavesdropper reveals himself by the process of his attack. Eavesdropping introduces inevitable errors to the exchanged quantum states that can later be detected by the communicating parties.
%Existing QKD protocols mainly differ by the way that quantum information is prepared and eavesdropping is detected.

%\subsubsection{Quantum States.}
At the core of every QKD protocol lies the exchange of quantum states.
In contrast to modern optical communication systems, where classical bits are encoded as an absent (0) or present (1) ``classical'' laser pulse in a certain time interval, QKD uses \emph{qubits} --- quantum objects, that can carry more than one bit of classical information at a time and exhibit a behavior that cannot be described within classical physics.
Very different physical systems can serve as qubits:
single photons, weak laser pulses, Fock states and squeezed states of light, half-spin quantum systems as trapped atoms and ions, or Rydberg atoms coupled to a cavity~\cite{Bouwmeester2000}.
%Here, a distinction between \emph{stationary} qubits used in quantum computers and \emph{flying} qubits used in QKD systems is made.
%\subsubsection{Information Encoding.}
Quantum information can be encoded using different types of \emph{observables}, i.e., physically measurable properties of qubits. For example, information can be encoded using polarization, phase, creation time of single photons, or quadrature, phase and amplitude of multi-photon coherent laser states~\cite{Bouwmeester2000,Scarani2009}.

\subsection{Common Functionality}
\label{sec-qkd-protocols}

%\subsubsection{Overview.}
We now sketch the functionality that is common to all QKD protocols discussed later. These protocols comprise a raw key distribution phase and a post-processing phase.

\paragraph{Raw key distribution.}
The first part of every QKD protocol establishes a raw secret key by
% which is then checked for errors that are introduced, for example, by eavesdropping.
transmission of qubits over a special \emph{quantum channel}. Ideally, such a channel should not alter the encoded information due to interaction of qubits with the transport medium (e.g. a change of polarization in a glass fiber). Distortions must be kept low in order to fulfill the requirements for a successful key distribution, because disturbances of the qubit states may have also been caused by an attacker.

%\subsubsection{Raw Key Exchange.}
During the raw key distribution phase, the communicating partners exchange qubits over the quantum channel.
Upon receiving a qubit, the recipient performs a measurement on some observable of the qubit and decodes a classical bit from its result according to a procedure determined by the chosen QKD protocol.
%However,  exchanging key bits via a quantum channel is not sufficient for generation of a secure key.
%Afterwards, the communicating parties have to consult with each other about their measurements over a classical authenticated channel --- an ordinary communication channel that delivers classical messages in an authenticated way.
Afterwards, the communicating partners consult about their measurements using a classical authenticated channel.
This procedure is specific to each QKD protocol and the result is a raw secret key.
If they deduce that an attacker might have disturbed the quantum information too severely, the key distribution has to be started over.

\paragraph{Post-processing.}
\label{sec-postprocessing}
After the raw key distribution phase, each of the communicating partners has obtained an individual raw key.
Perfectly correlated keys are very improbable, so error correction (e.g. \emph{low density parity check}~\cite{Gallager1962}, \emph{cascade}~\cite{Brassard1993}, or \emph{polar codes}~\cite{Jouguet2012}) has to be performed.
%These algorithms are also used in a modern standard communication and broadcasting, e.g. low density parity check in DVB-T2 \cite{}.
Afterwards, privacy amplification is applied to generate the final key from the error-corrected raw key.
This ensures security even against an eavesdropper that may have observed a small number of bits undetected during the raw key exchange or the error correction.
The resulting secret key can then be used as a key for OTP or Advanced Encryption Standard (AES) encryption.
%The encrypted data is then communicated over a classical channel. The quantum channel is used only for secure key distribution.

As described above, QKD requires an authenticated classical channel between the communicating partners.
Such an authenticated channel can be established using a short pre-shared secret or by relying on a typical TLS connection.
Recently, it was proposed to realize authenticated channels based on laws of quantum physics~\cite{Goorden2013,Nikolopoulos2017}.
We remark that the authenticated channel used in a QKD protocol needs to remain secure only while the QKD protocol is executed.

\subsection{Protocol Families}

There are many different ways QKD protocols are implemented.
For our analysis, we categorize them by the way information is prepared (prepare-and-measure or entanglement based) and by the type of variables (discrete variables (DV), continuous variables (CV), or distributed phase reference (DPR)). %or by the alignment scheme (one-way or plug\&play).

\subsubsection{Classification by information preparation method.}
We describe categories for QKD protocols based on how the quantum states are prepared.
%\subsubsection{Classification by Setup Type.}
%QKD implementations can be categorized by the type of setup used.

\begin{figure}[!ht]
\centering
 \subfloat[]{\includegraphics[width=0.5\textwidth]{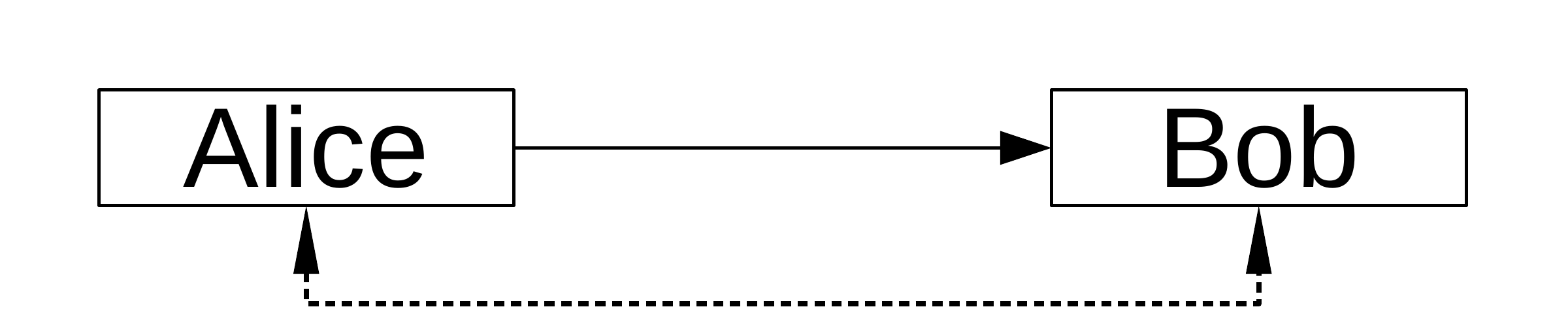}\label{prepare-and-measure-fig}}\hfill
 \subfloat[]{\includegraphics[width=0.5\textwidth]{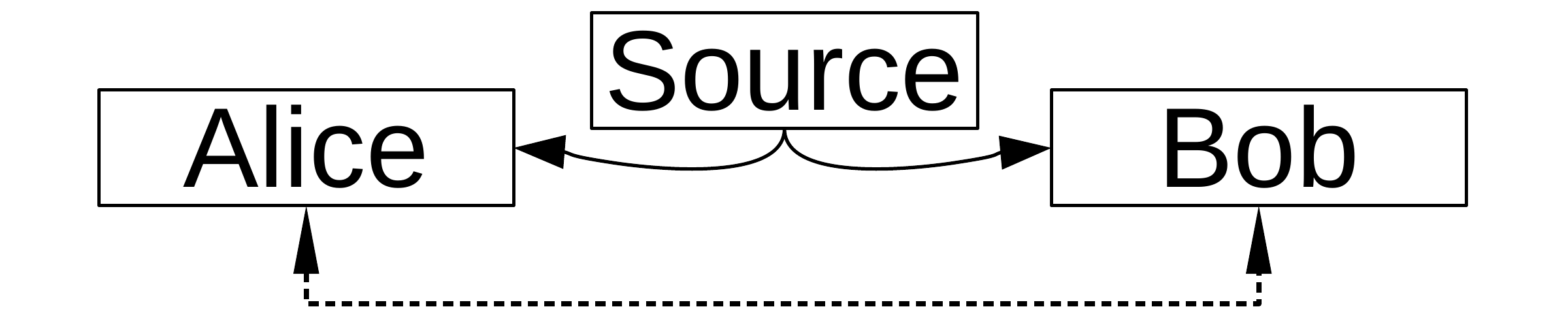}\label{entanglement-based-fig}}
\caption{(a) Prepare-and-measure protocol (e.g. BB84~\cite{Bennett1984}), (b) Entanglement-based protocol (e.g. E91~\cite{Ekert1991}). Solid line denotes the quantum channel, dashed line stands for the authenticated classical channel. Arrows denote the direction of information flow.}
\label{protocol-types-fig}
\end{figure}

\paragraph{Prepare-and-measure.}
In prepare-and-measure protocols (Fig.~\ref{prepare-and-measure-fig}) a sender Alice actively prepares an information carrier, encodes information within it and sends it to one or more recipients. 
Prominent representatives of this protocol category are the protocol developed by Bennett and Brassard (BB84)~\cite{Bennett1984} or derived protocols, such as~\cite{Scarani2002,Stucki2005}.

\paragraph{Entanglement-based.}
Entanglement-based protocols (Fig.~\ref{entanglement-based-fig}) involve a source producing \emph{entangled} particles --- multiple quantum objects that can be described by a correlated quantum state violating local realism~\cite{Bell1964}.
%Such objects can be considered as a single object regardless the distance between them.
A measurement on some observable of one of the objects instantly affects the state of the other object. This can be observed by \emph{Bell tests}~\cite{Bell1964} --- a procedure allowing the verification of the useful entanglement~\cite{Scarani2001} of the initial particles. 
The qubits are detected by the communicating parties, and, because of the non-classical correlations between these particles, Alice and Bob can share a (quantum) secret without direct exchange of information.
A prominent representative of this protocol category is the E91 protocol, developed by Ekert~\cite{Ekert1991}.

 %Stabilization of the set-up is often required if information is encoded in a observable that is sensitive to noisy environment. In the case of the one-way configuration the source is at Alice's set-up. During the propagation over the quantum channel noise  to Bob and the set-up should be stabilized passively. In plug\&play systems Bob has the strong light source and sends

%\todo[inline]{Beschreibe high-level, was der Unterschied zwischen den einzelnen types of observables (was bedeutet eigentlich discrete und continuous variable? Was ist der Unterschied? Was ist der Unterschied zu distributed phase reference)? Unterscheidet sich das Setup? Gibt es Protokolle, die die Verfahren gemeinsam benutzen? Sagen, dass die in den folgenden Kapiteln genauer erklaert werden.  }

\subsubsection{Classification by the type of variables.}
QKD protocols can also be classified by the type of variables used for the information carriers.
%Here, different classes of protocol require fundamentally different device setups.
%Classification of the QKD protocols by the types of observables is more constructive, since different families of protocols possess fundamentally different information carriers and, thus, set-ups. While some protocols can be implemented both prepare-and-measure and entanglement based variants of the set-up, the observables used for information encryption are a crucial part of an protocol. In following, we give a short description for discrete variable, continuous variable and differential phase reference protocols.

\paragraph{Discrete variables.}
For the protocols with discrete variables (DV), the values of the information carrying observables are discrete. Most commonly, qubits are transmitted using single photons or weak laser pulses.
%Here, the average number of photons transmitted per pulse is $N=1$.
In principal, half-spin particles (e.g. electrons) can also be used, but the transmission of such particles is problematic.
The information can be encoded, for example, in time, polarization, spin, or phase. The source can be implemented as a prepare-and-measure system or as an entanglement-based system.
DV protocols require expensive and inefficient single-photon source and detector devices.
%The equivalence \added{of the security proofs} of these two types for some protocols has been shown in~\cite{Bennett1992}.\todo{for O,A: why do you mention this here? is this relevant here? if yes, please formulate more concretely what this means. if no, remove this sentence.}
%\deleted{In all cases the communication parties are connected by classical channels, additionally to a quantum channel. The classical channel \todo[inline]{allows to ...? and }needs to be authenticated at the beginning of key distribution session.}
%\todo[inline]{Gibt es eine einheitliche Begruendung, warum diese Verfahren sicher sind?}\oleg{Nein, die Sicherheit von jedem Protokoll muss gesondert betrachtet werden. Im Allgemeinen, ist QKD sicher, weil ein Angreifer sich verrät, wie schon am Anfang des Unterkapitels gesagt wurde}
Prominent representatives of this protocol category are~\cite{Bennett1984,Bennett1992,Ekert1991,Scarani2002}.

\paragraph{Continuous variables.}
Continuous variable (CV) protocols are an alternative to DV protocols that, instead of qubits (e.g. single photons and weak laser pulses), use many particle states (e.g. squeezed or coherent states of light).
Hereby no discrete variables are detected --- zeros and ones --- but the observation of the continuous spectrum of the quadrature components of light is performed, e.g., by homodyning techniques~\cite{Diamanti2015}.
%For example, by using quadratures $q$ and $p$ of coherent states and their superpositions, see \cite{Diamanti2015}.
Detection of quantum states is also performed differently in CV protocols compared to DV protocols.
Here, standard components for quantum communication are used. For instance, homodyne or heterodyne detection schemes~\cite{Diamanti2015} (In homodyne schemes, the signal and a local oscillator have identical
frequencies. In heterodyne schemes, these frequencies differ) are employed. This is much faster and more efficient than the detection of single photons. Most of these protocols can be implemented both as prepare-and-measure as well as entanglement-based schemes. 
Prominent representatives of this protocol category are~\cite{Grosshans2002,Ralph1999}.

\paragraph{Distributed phase reference.}
A third family of QKD protocols --- distributed phase reference (DPR) protocols --- uses discrete variables for encoding of information, but at the same time the security is guaranteed by observing the coherence of subsequent pulses.
Bits may be encoded in a sequence of pairs of pulses~\cite{Stucki2005} or in the phase of subsequent pulses~\cite{Inoue2003}.
The two approaches may also be combined into a two dimensional QKD protocol~\cite{Bacco2016}, where several bits can be encoded by two subsequent pulses.
%\deleted{While unconditional security has been shown for the discrete variable protocols for the distributed phase reference protocols this still has to be proven.}
DPR protocols require similar devices as DV protocols, namely, single photon sources and detectors.
Prominent representatives of this protocol category are~\cite{Bacco2016,Inoue2003,Stucki2005}.

\subsection{Existing Protocols}

%\todo[inline]{Todo for Oleg: Give overview of most relevant protocols (BB84,E91,"DPS protocol","CV protocol").Say for each protocol what type of setup and what type of observable. Do not explain in detail.Maybe cite other protocols (e.g. SARG04,BBM92,\ldots) if relevant.Max 3/4 page.}
%\todo[inline]{Viele Protokolle haben prepare-and-measure und entanglement based Varianten, die auch sehr oft equivalent sind. Wie BB84 und BBM92, bzw. Gaussian modulated CV QKD, cites!}

%In the previous section we introduced two ways of classification of the QKD protocols.
%The equivalence between these two implementations for Gaussian modulated protocols has been shown in~\cite{Grosshans2003}.
In the following, we briefly summarize the functionality of a few prominent QKD protocols and assign them to the categories presented above.

%\todo{Bezug auf 'Type of Information Preparation' and 'Type of Observable'}

\paragraph{BB84.}

The first QKD protocol was proposed by Bennett and Brassard in 1984 (BB84)~\cite{Bennett1984}. This protocol is a prepare-and-measure protocol, belongs to the DV protocol family, and uses discrete states of photons for information encoding. The information can be carried by polarization or by the phase of single photons. 

\begin{figure}[!ht]
\centering
 \includegraphics[width=0.8\textwidth]{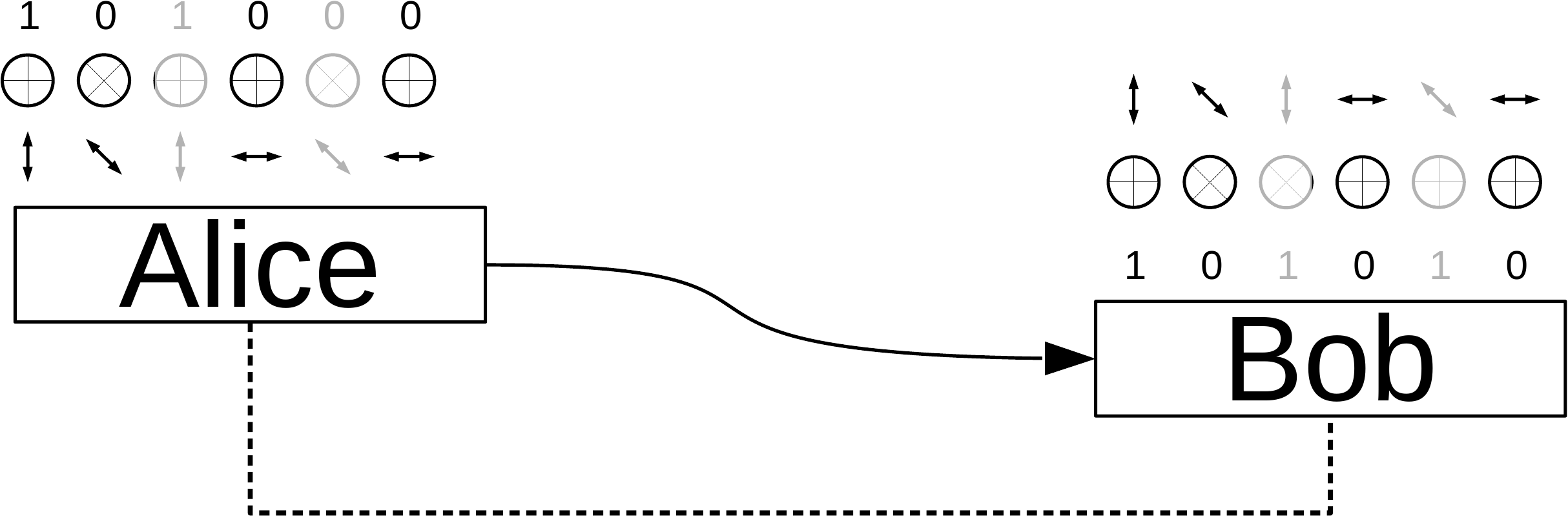}
\caption{BB84 protocol.}
\label{fig-BB84}
\end{figure}

%Since the introduction, security of this protocol has often been treated \cite{}, several different attacks on the experimental realizations have been carried out \cite{}. %And therefore different variants of BB84 have been proposed. Either one could add so-called decoy states \cite{} or a different encoding method and obtain a new protocol SARG04 \cite{Scarani2002}. 
The BB84 protocol with polarization encoding works as follows (see Fig. \ref{fig-BB84}): 
Alice chooses randomly one bit --- 0 or 1, and one of two polarization bases: rectilinear, denoted as $\bigoplus$, or $\pm45^{\circ}$, denoted as $\bigotimes$. Then, Alice encodes that bit within the chosen basis and sends it to Bob.
Bob also chooses a basis randomly and independently from Alice, in which he detects the photon.
Statistically, in half of all cases his choice does not correspond to the basis chosen by Alice.
If Bob chooses a wrong basis, he can measure the correct bit with a probability of 50\%, so an error rate of 25\% is introduced.
To get rid of these errors Alice and Bob exchange publicly over an authenticated classical channel information about their chosen bases and dismiss all of the events detected in the wrong bases. This procedure is called \emph{key sifting}. Afterwards, parts of the sifted key are exchanged between Alice and Bob, compared, and their quantum bit error rate (QBER) is estimated.
If the QBER does not exceed 20\%~\cite{Chau2002}, a secure connection can be established using two-way error correction and privacy amplification.
For this, post-processing algorithms are used as described in Sec.~\ref{sec-postprocessing} and a secret key is distilled.

\paragraph{E91.}
Another milestone in DV QKD protocols is the protocol proposed by Ekert in 1991 (E91)~\cite{Ekert1991}. It was the first entanglement-based protocol that does not require the source to be a part of either Alice's or Bob's setup. 
A source of entangled photons distributes distinct photons of a qubit pair to the different communicating partners Alice and Bob. Then, Alice and Bob choose a detection basis randomly and independently, perform measurements, and obtain a raw key. With a random subset of this raw key, they test for a violation of Bell's inequalities. If the test succeeds, Alice and Bob start key sifting, error correction, and privacy amplification procedures as in BB84. Otherwise, they have to start the key distribution over again.

\begin{figure}[!ht]
\centering
% \subfigure[]{\includegraphics[width=0.49\textwidth]{Figures/BB84.png}\label{fig-BB84}}\hfill
 \includegraphics[width=.8\textwidth]{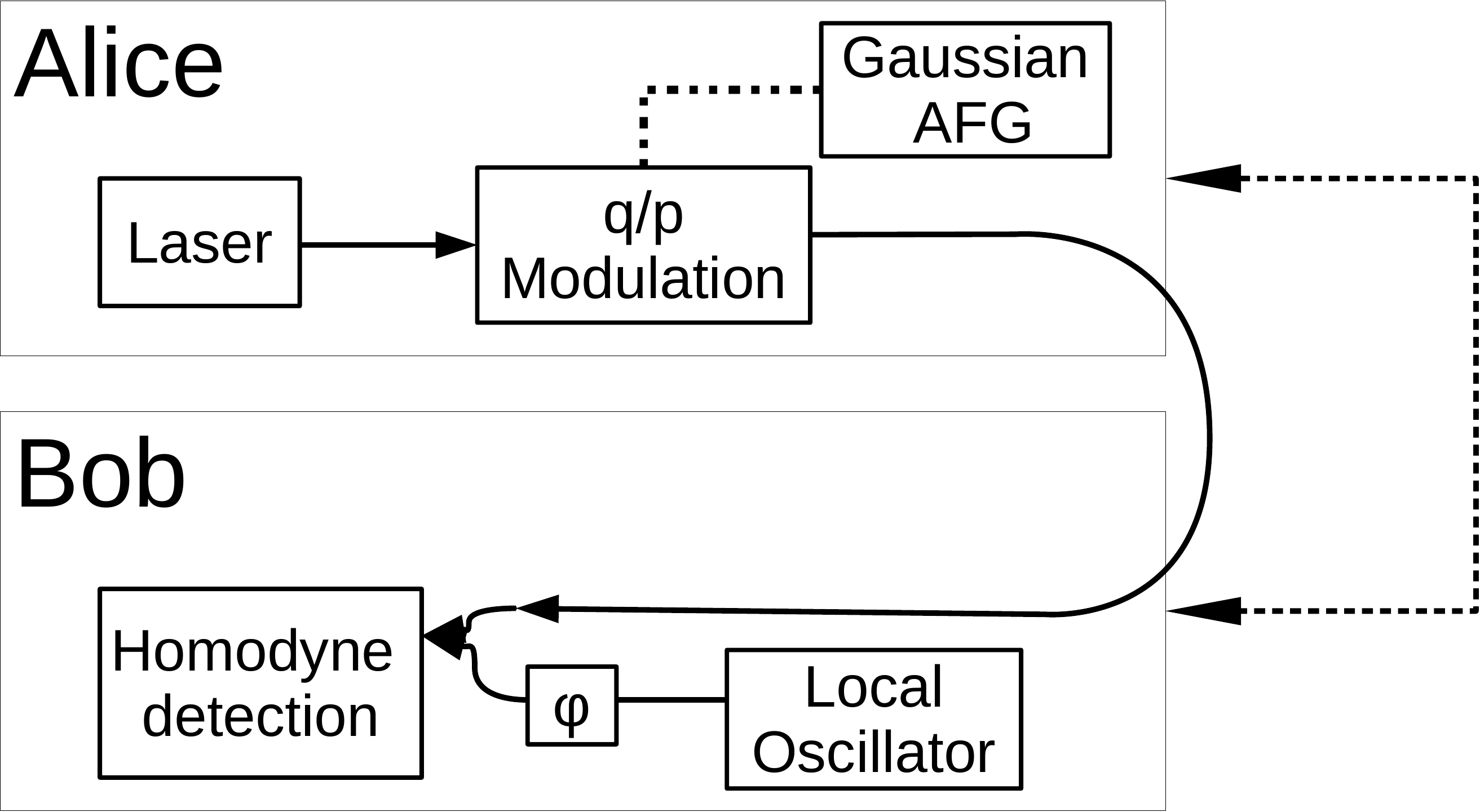}
\caption{Schematic of a CV protocol, as in~\cite{Grosshans2002} with homodyne detection.
AFG is an Arbitrary Function Generator.}
\label{fig-CV}
\end{figure}

\paragraph{GG02.}
In Fig.~\ref{fig-CV} the CV protocol implemented by Grosshans and Grangier is shown~\cite{Grosshans2002}. Here, a single-mode coherent state produced by a laser at the wavelength of \SI{1550}{\nano\meter} is modulated in one of its quadratures $p$ or $q$ by using a random, Gaussian distributed modulation. Then the signal is transmitted over a noisy channel to Bob, who measures only one of the quadratures using homodyne techniques. Therefore, a local oscillator (e.g. laser light) has to be implemented locally by Bob or transmitted over the quantum channel from Alice. The choice of the quadrature is performed by the addition of the phase $\varphi\in\{0,\pi/2\}$ randomly to the local oscillator. Information reconciliation and error correction are performed after the detection.

\subsection{Performance}
The aforementioned QKD protocols can be run over free space or via glass fibers. Depending on the communication medium, different secret key generation rates and effective distances are observed. Typical key rates for DV protocols are up to several \si{\kilo\bit\per\second} on the distance of several \SI{10}{\kilo\meter} and up to several \si{\bit\per\second} over approximately \SI{100}{\kilo\meter} distance (cf. Tab.~\ref{fig-performance}) via fiber cables. Recently, DV-based quantum key distribution via satellite has been demonstrated over a distance of \SI{1203}{\kilo\meter} at a key rate of \SI{1}{\kilo\bit\per\second} \cite{Liao2017}.
For CV protocols the key rate is comparable --- up to \SI{10}{\kilo\bit\per\second} for channels of a few \si{\kilo\meter} and up to \SI{150}{\kilo\meter} effective distance. %The COW protocol has achieved a record length --- 307 km with a key rate of 3.18 bit/s (12.7 kbit/s at 104 km length). Entanglement based communication --- 144 km \cite{Ursin2007}.
In Tab.~\ref{fig-performance}, performance measures for fiber based QKD protocols are listed. For all QKD protocols, the key rate decreases exponentially with the communication distance due to noise and losses in the quantum channel.
It has been proven that the maximal key rate depends solely on these losses~\cite{Takeoka2014,Pirandola2017}.
Lost information is assumed to be gained by a potential attacker in the case of prepare-and-measure protocols, resulting in a failed protocol execution. In the case of entanglement-based protocols, this reduces the violation of the Bell inequality, ultimately leading to a failed Bell test and protocol execution.

%DPR protocols: Gisin (2009): 250 km - 15 bit/s or 100km - 6kbit/s. Tokyo - up to 2 kbit/s secret key.

\begin{table}
\caption{Performance of different protocols. PaM: prepare-and-mesure protocols, EB: entanglement-based protocols.}
\label{fig-performance}

\centering
%\resizebox{\columnwidth}{!}{
\setlength{\tabcolsep}{.5em}
\bgroup
\def\arraystretch{1.1}
\begin{tabular}{l c c c c }
\noalign{\smallskip}
\hline
\noalign{\smallskip}
% at \SI{100}{\kilo\meter}}	&	\textbf{Key rate\\at max. distance}
\textbf{Protocol}	&			\textbf{Type}	&	\textbf{Source} & \textbf{Key rate}	& \textbf{Max. dist.}	 \\
	&				&		&	at \SI{100}{\kilo\meter} &   \\

\noalign{\smallskip}
\hline
\noalign{\smallskip}

Yin \cite{Yin2016} (2016)	&			DV	&	EB	&	\SI{2}{\kilo\bit\per\second}	&	 \SI{404}{\kilo\meter}	\\
Korzh \cite{Korzh2015} (2015)	&			DPR	&	PaM	&	\SI{10}{\kilo\bit\per\second}	&	\SI{307}{\kilo\meter}	\\
Wang \cite{Wang2012} (2012)	&			DPR	&	PaM	&	\SI{20}{\kilo\bit\per\second}	&	 \SI{260}{\kilo\meter}	\\
Stucki \cite{Stucki2009}  (2009)	&			DPR	&	PaM	&	\SI{6}{\kilo\bit\per\second}	&	 \SI{250}{\kilo\meter}	\\
Liu \cite{Liu2010} (2010)	&			DV	&	PaM	& \SI{15}{\bit\per\second} (\SI{200}{\kilo\meter}) &	 \SI{200}{\kilo\meter}	\\
Huang \cite{Huang2016} (2016)	&			CV	&	PaM	&	\SI{500}{\bit\per\second}	&	 \SI{100}{\kilo\meter} \\
Honjo \cite{Honjo2008} (2007)	&			DV	&	EB	&	\SI{0.59}{\bit\per\second}	&	 \SI{100}{\kilo\meter}	\\
Jouguet \cite{jouguet2013} (2013)	&			CV	&	EB	& \SI{200}{\bit\per\second} (\SI{80}{\kilo\meter}) & 	\SI{80}{\kilo\meter}  \\

\end{tabular}
\egroup
%}

\end{table}

Besides the key rate and distance, the compatibility of the system with the existing communication infrastructure is important.
For example, DV QKD protocols require quite expensive single photon detectors, single- or entangled-photon sources and precise time measuring devices. Simultaneously, the typical distribution distances and rates for the secret key distribution allow for use only in metropolitan network areas. Imperfections in the single-photon sources make photon number splitting attacks possible (see Sec.~\ref{sec.attacksqke}).

CV protocols are a more recent class of protocols that offer higher secret key rates and lower costs for implementation, because neither single photon sources nor single photon detectors are required. Standard components for optical communication can be used.
A recent experiment showed that CV protocols can be applied even in a geostationary satellite for standard optical communication achieving much longer communication distances~\cite{Guenthner2017}.
However, the security of CV protocols against side-channel attacks is not as well understood as for DV protocols~\cite{Diamanti2015,Laudenbach2017} (cf.\ Sec.~\ref{sec.attacksqke}).

DPR protocols currently achieve the best performance results (Fig.~\ref{fig-performance}). Furthermore, multi-dimensional QKD schemes, such as DPR protocols, allow to transmit more than one bit of classical information in a single qubit~\cite{Bacco2016}.

%!TEX root = main.tex

\subsection{Security}
\label{sec.attacksqke}

A QKD protocol is considered secure if, after the protocol execution, the communication partners Alice and Bob know a common secret key and an eavesdropper Eve has not obtained any information about the key.
We now summarize work analyzing the security of QKD protocols and discuss theoretical and practical attacks on implementations of QKD.

%\todo[inline,color=red]{for Oleg: General structure for this subsection: first mention initial assumptions about device setup of Alice and Bob (e.g., a perfect single-photon source, idealized source, detectors, noiseless channels). then say that these assumptions do not hold in practice. describe how assumptions are modified and whether security can still be proven.}
%\todo[inline,color=red]{for Oleg: check if a non-quantum physicist (e.g., computer scientist) can understand this!}

\subsubsection{Security models and proofs.}
When analyzing the security of a QKD protocol the goal is to show security against a powerful attacker, Eve, that potentially possesses perfect technology.
For example, Eve may be able to extract and store qubits for an arbitrary duration and perform any quantum operation or measurement on them. However, according to fundamental quantum physical laws, Eve can neither clone nor measure the state of the system perfectly and resend a new particle without leaving a trace.
%All noise and loss effects of imperfect hardware are assigned to Eve's interceptions.
In addition, usually the existence of an authenticated classical channel between the communication partners or a short pre-shared key is assumed. This is necessary to guarantee data integrity and authenticity, so that Eve cannot perform an impersonation attack or change the classical data sent.
We stress that the authenticated channel does not need to provide any confidentiality guarantees.
%In this setting, unconditional information-theoretic security of the protocol against general attacks can be assessed under the assumption of trusted setups.

%The attacks on QKD can be divided in several types, that are described below.
%\subsubsection{Idealized attacks.}
%\label{sec-thAttacks}
An attack on a QKD system is called \emph{individual} if Eve measures each qubit separately.
In a \emph{collective} attack, Eve still interacts with each qubit separately, but she
may measure all the auxiliary systems used for the interactions jointly.
If Eve is allowed to attack several sent qubits simultaneously, the attack is called \emph{coherent}.
Renner et al.\ \cite{Renner2005} prove the security of a wide range of QKD protocols against coherent attacks.

QKD security proofs rely on information theory and do not depend on computational hardness assumptions. This fundamental difference in comparison to currently used key distribution methods guarantees the long-term security of QKD.
However, idealized assumptions in QKD security proofs lead to incomplete security models. For realistic security guarantees about actual implementations, more assumptions regarding hardware and software are required. Indeed, attacks exploiting imperfect devices and insecure software may be possible, as we describe below.

Depending on protocol families, proven security guarantees against theoretical attacks vary.
While some DV protocols have been shown to be unconditionally secure~\cite{Lo1999,Shor2000}, similar proofs for CV and DPR protocols are still missing. An overview of security proofs for CV protocols is given by Diamanti, Kogias, Laudenbach and others~\cite{Diamanti2015,Kogias2017,Laudenbach2017}. 
A security analysis of DPR protocols is provided by Moroder et al.~\cite{Moroder2012}.

As an example, we discuss the security of BB84 against an \textit{intercept-resend attack}, which is a special case of an individual attack.
In this attack, Eve chooses a basis randomly and detects the state of particles. She has a probability of 50\% to choose a wrong basis. Afterwards, she prepares a replacement for the detected qubit and sends it to Bob. In that way, she induces a 25\% QBER in Bob's key.
However, as shown by Shor and Preskill, Alice and Bob know the key distribution session might have been compromised~\cite{Shor2000} if the QBER exceeds 11\%. Other strategies, for example, detection of not every qubit or detection using an intermediate basis are disadvantageous for Eve, since she obtains less information about the secret key.
In the case of entanglement-based protocols, during the measurement of qubits, Eve destroys the nonclassical correlations between the particles, so a Bell test during the key processing fails.
In summary, security proofs for QKD protocols show that an attacker reveals himself when trying to eavesdrop on the quantum states sent over the network.
This is what makes QKD so powerful in comparison to classical key distribution.

%During joint attacks Eve is assumed to investigate several qubits simultaneously. And collective attacks is a intermediate category between individual and joint attacks. Eve is assumed to address each qubit, but is also able to detect several qubits coherently. For these attack types the same bound of QBER<11\% as security condition of the protocol is found \cite{}.

\subsubsection{Attacks on implementations.}
\label{sec-prAttacks}

Even for protocols that have been proven unconditionally secure, side-channels and non-perfect setups can lead to weaknesses. Every single implementation then requires a special treatment. As for classical cryptography, the developers of QKD systems search for any kind of weakness in their implementation. Security proofs have to be adapted and side-channels must to be analyzed. 

For example, the creation of tailored single photons is non-trivial. In most cases, there is a non-negligible probability for pulses with a photon number larger than $1$. Thus, if there is more than one photon in a weak laser pulse, Eve can pick some photons with a beam splitter and gain information without being noticed. This type of attack is called a \textit{photon number splitting attack}. As a countermeasure, protocols have been modified: decoy states have been added to BB84~\cite{Lo2005} and new protocols, such as SARG04, have been developed~\cite{Scarani2002}. 

Hijacking a quantum channel by a \textit{trojan horse attack}, information about Alice's and Bob's setups can be extracted~\cite{Gisin2006} or even manipulated~\cite{Lydersen2010}. For example, if Eve obtains information about Alice's choice of bases in real-time, she can perform a successful intercept-resend attack as she is no longer limited to guessing the bases randomly. 

Another possibility is bright illumination of Bob's detectors via the quantum channel. This can allow the attacker to control the measurement results of Bob. Lydersen et al. describe how an attacker could successfully obtain the complete secret key and remain unnoticed~\cite{Lydersen2010}.

Crucially, all these attacks must be performed physically and during the actual key distribution. This is a fundamental difference to classical key distribution protocols, whose security might be broken by attacks that were unknown at the time of the distribution.

%\subsubsection{Intercept-resend.}
%In the case of individual attack single qubits are under observation. The easiest scenario is the \textit{intercept-resend attack}. Hereby, Eve chooses a basis randomly and detects the state of particles. She has a probability of 50\% to chose a wrong basis. Afterwards she prepares a replacement for the detected qubit and sends it to Bob. In that way she induces 25\% BER in Bob's key. If the QBER exceeds 11\%, the key distribution session has been compromised \cite{}. Other strategies, as for example detection of not every qubit or detection using a intermediate basis are disadvantageous for Eve, since she obtains less information about the secret key.

%In the case of entanglement-based protocols, during the measurement of qubits, Eve destroys the nonclassical correlations between the particles, so a Bell test during the key processing fails.

\subsubsection{Device-independent QKD.}
Device-Independent QKD is an approach aiming to dispense with the assumption of trust in the own setup hardware~\cite{Mayers1998}. Hereby, the security of the whole QKD system should be evaluated by a quantum-correlation test, i.e., a Bell test, similar to the E91 protocol~\cite{Vazirani2014}. Since purely device-independent protocols are hard to realize, measurement-device-independent QKD protocols have been developed~\cite{Lo2012,Tang2014,Yin2016}. 

%\subsubsection{Experimental Attacks}

%\paragraph{Trojan horse attacks}
%This category of attacks contain attacks collecting information about the settings of Alice and Bob module or even manipulating both devices using only the quantum channel for access \cite{}.
%One of most famous attack on a QKD system was performed by .

%\paragraph{Side-channel attacks}
%There are plenty of QKD set-ups. Each set-up can be attacked on its special way. A possible solution for that is a device independent QKD. For DV protocols a loophole free Bell test could be a necessary condition for it. A measurement device independent protocol is available~\ref{MDI-CV-sec} and requires weaker assumptions.

%\todo{for O,A: Tabelle, welche Angriffe bei welchen Verfahren funktionieren?}

%!TEX root = main.tex

\section{Quantum-based Key Distribution in Multi-User Networks}
\label{sec.qkehubs}

%\oleginline{Es gibt eine Verbindung zwischen der Sicherheit des zweiparteien Protocols und eines N-Parteien Protocols \cite{Scarani2001}. Das heißt unser Hub wäre sicher bei einer echetn vier Parteien kommunikation. Das, was unser Projekt macht ist eine zwei Parteien Kommunikation zwischn zwei beliebigen Partnern. Daher trifft hier immer noch der Beweis der bipartiten Kommunikation.}

So far, we have discussed QKD in a setting where two communicating partners are connected directly by a dedicated quantum channel.
However, in reality dedicated communication channels usually do not exist between every two communicating partners.
Instead, communicating partners are connected via hubs through which the communication is routed.
To realize long-term secure communication using QKD technology, hubs need to be developed that are compatible with QKD.
%This requires the development of a hub technology which is compatible with QKD protocols.
%This requires the development of hubs for establishing connections between users that are not directly connected with a network link.
In the following, we first describe how the Internet is organized using hubs and thereby identify requirements for QKD protocols that should be used in such networks.
Then, we discuss two approaches for realizing QKD hubs which allow for using QKD in a hub-based network.

\subsection{Network Characteristics of the Internet}
\label{sec.infrareq}

%We briefly describe the functionality of the Internet and thereby identify properties that are relevant for designing suitable key exchange technology.

\begin{figure}
\centering
\includegraphics[width=.8\textwidth]{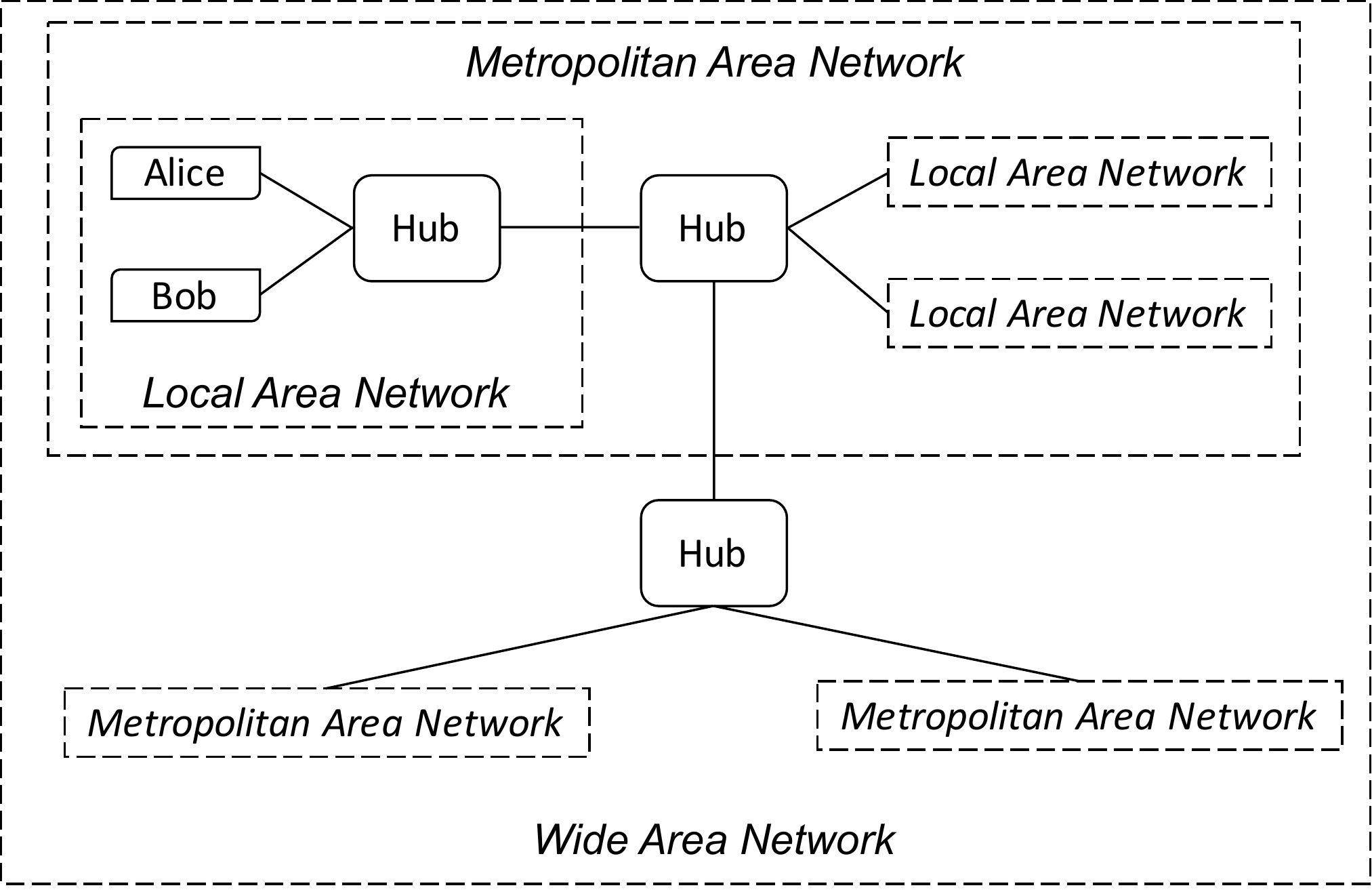}
\caption{The various network layers of the Internet that are connected via hubs.}
\label{fig.internettopology}
\end{figure}

The Internet consists of an extensive number of different types of devices that potentially communicate with each other.
These devices may be small (e.g. mobile phones, wearables) or large (e.g. desktop computers).
Communication should be possible between any two devices and the devices should be able to join and leave the network dynamically.
Distances between the devices may be very long (e.g. several thousand of kilometers) and many services run on the Internet require low latency (e.g. less than a second) and high transfer rates (e.g. several megabits per second).
These properties are achieved on the Internet by using a multi-layer network topology consisting of local area networks, metropolitan area networks, and wide area networks.
The different network layers are connected via hubs (Fig.~\ref{fig.internettopology}).
Messages exchanged between clients are routed through these hubs.
Secure end-to-end communication is usually achieved by establishing a TLS connection between any two end points.

\subsection{Trusted Node-based Hubs}

\begin{figure}[ht]
  \centering
   \includegraphics[width=0.6\textwidth]{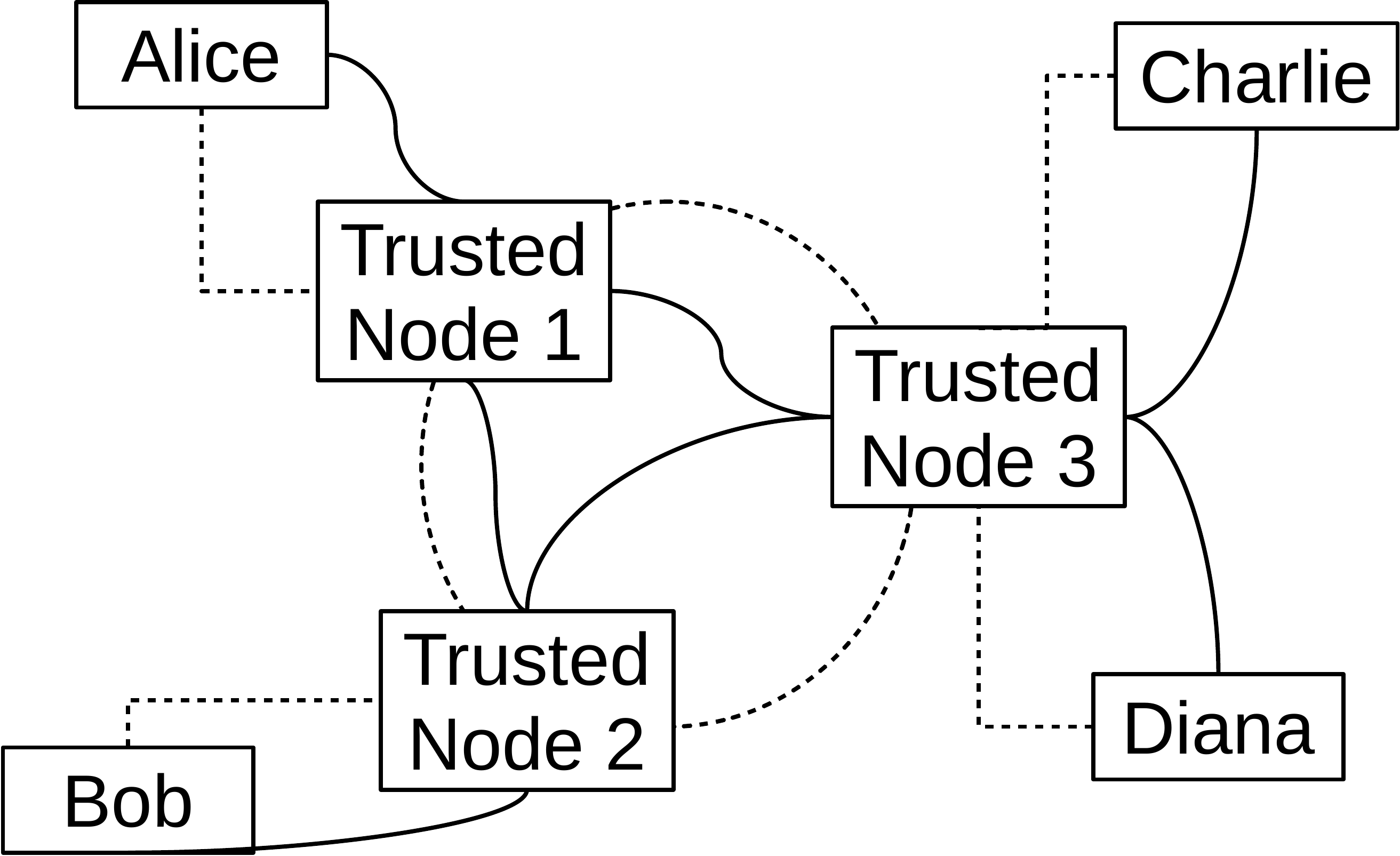}
   \caption{Scheme of a QKD network link with trusted nodes. Solid lines are quantum channels, dashed lines denote classical channels.} 
   \label{fig-trusted-nodes}
   \end{figure}

Network hubs for QKD can be realized using trusted nodes.
In this approach, key material is not directly exchanged between the communicating partners but relayed over the trusted nodes.
Here, each of the communicating partners, Alice and Bob, are connected to a trusted node in close distance and the trusted nodes are also connected (Fig.~\ref{fig-trusted-nodes}).
In order for Alice and Bob to establish a secure connection, they exchange keys with their trusted nodes, and the trusted nodes also exchange keys between themselves.
Then, an end user key for Alice and Bob is computed from the key material generated on the communication path.
This method for key distribution in distributed networks only provides security as long as none of the trusted nodes on the communication path is compromised~\cite{Elkouss2013}.

%QKD networks based on trusted nodes do not support end-to-end security. Keys are only exchanged between neighboring trusted nodes (see Fig.~\ref{fig-trusted-nodes}). 
%Then, an end user key is computed for the communication parties from the key material generated by the trusted nodes on the communication path.
%Secure communication is only guaranteed as long as the key relay nodes do not leak any key material.

On the positive side, such hubs allow to extend the distance limitations imposed by QKD protocols.
Assuming there exist sufficiently many trusted nodes between the communicating parties, keys can be relayed several times and QKD distance limitations only account for each key relay individually.
On the negative side, we have seen that trusted node-based hubs do not support end-to-end security because the quantum states are destroyed in each hub.

The feasibility of building a trusted node based QKD network has been shown in several field tests: DARPA network in 2005~\cite{Elliott2005}, in Austria in 2008~\cite{Peev2009}, in Japan in 2011~\cite{Sasaki2011} and China~\cite{Wang2014}. Currently, a 2000 km long link connecting Beijing to Jinan, Hefei and Shanghai is being installed~\cite{china2017}.
%http://spectrum.ieee.org/telecom/security/chinas-2000km-quantum-link-is-almost-complete

\subsection{Quantum Hubs}
An alternative approach for building QKD networks is to use quantum hubs~\cite{Herbauts2013}.
Here, the quantum information carriers are routed from Alice to Bob directly without any detection or optical-electrical-optical conversions in between. One can either use active optical switches~\cite{Chen2010,Chang2016} or wavelength division multiplexing (WDM) to address different recipients. Recently, several experiments showing entanglement distribution via WDM in glass fiber at telecommunication wavelengths have been reported~\cite{Ghalbouni2013,Cao2015}.
%Another promising research direction for QKD hubs is the development of hubs whose security is based on the same laws of quantum physics as the corresponding QKD protocols.Here, the quantum information carriers are routed from Alice to Bob directly without any detection or optical-electrical-optical conversions in between.

\begin{figure}[ht]
  \centering
   \includegraphics[width=0.5\textwidth]{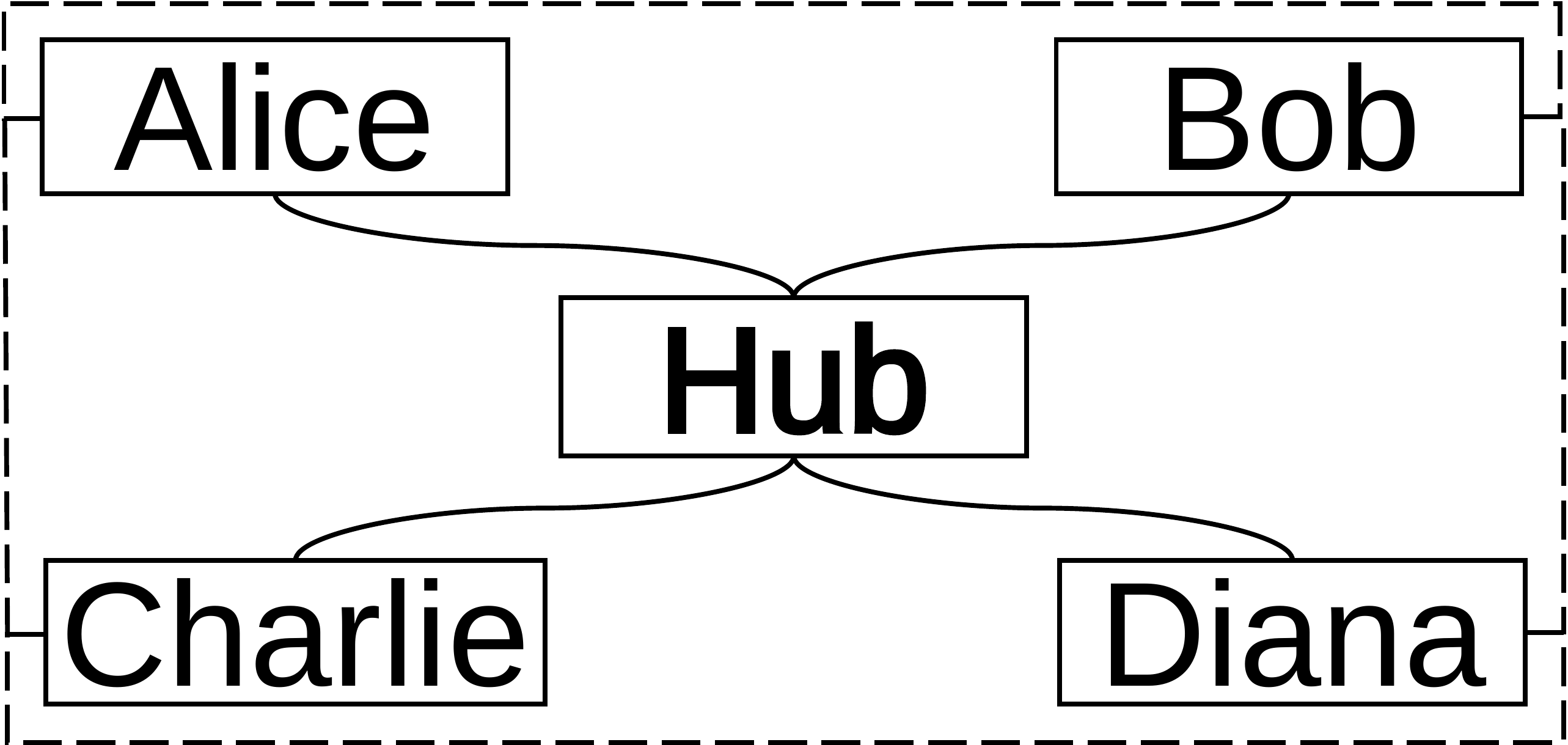}
   \caption{Scheme of a star-shaped QKD network. Solid lines are quantum channels, dashed lines are classical channels.}
   \label{fig-qhub}
  \end{figure}

A novel approach for such all-optical network is an extension of entanglement-based DV protocols. 
Such a quantum hub works for entanglement based protocols (e.g. E91 or BBM92) and is depicted in Fig.~\ref{fig-qhub}.
The photons are created in the quantum hub at distinct wavelengths and then distributed to different end users by using standard wavelength division multiplexing techniques.
Such quantum hubs allow for real end-to-end security. 
%The security is equivalent to that of two party communication with these protocols.
%\todo{Todo for Oleg: update figure for all optical network}
%\todo{Security of quantum hubs?}
%\todo{for O,A: please explain in more detail. for which protocols can we realize this?}

\begin{figure}[ht]
\centering
\includegraphics[width=0.5\textwidth]{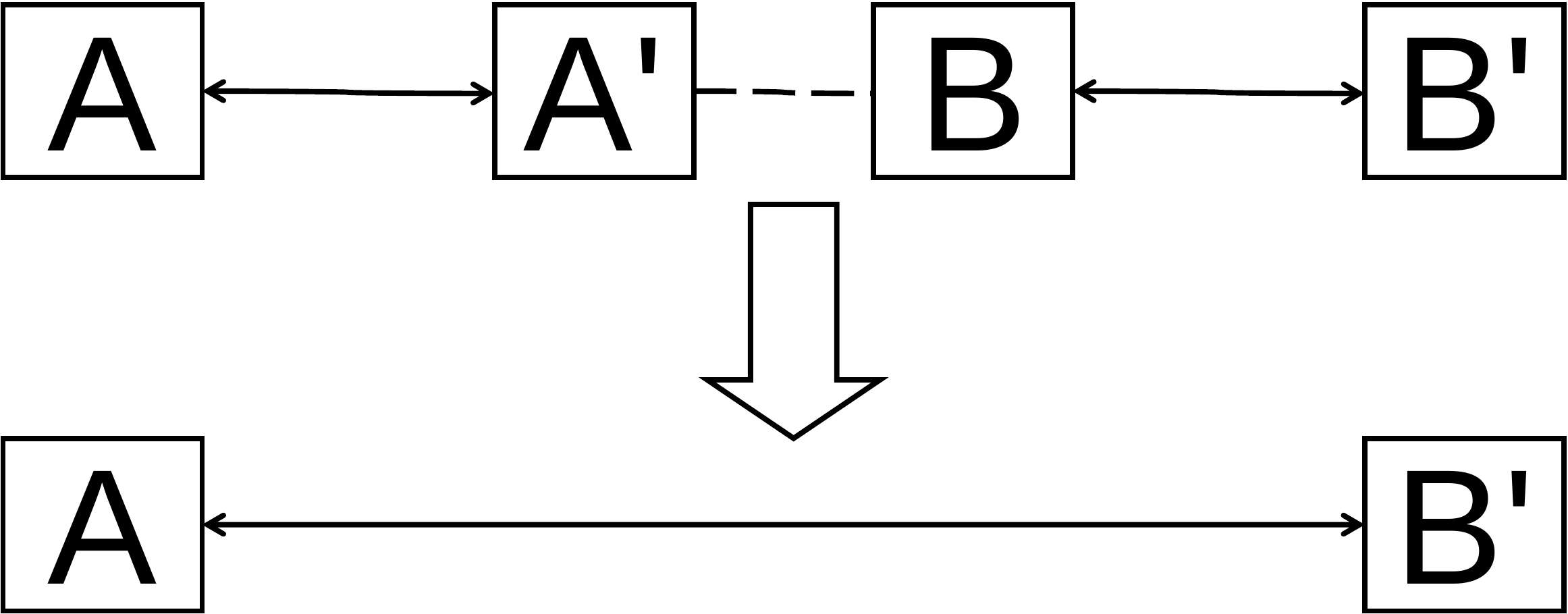}
\caption{Schematic structure of a quantum repeater.}
\label{QRep-fig}
\end{figure}

%\subsection{Quantum Repeater}
\label{QRep-sec}
Quantum hubs have similar distance limitations as the corresponding QKD protocols.
%, simultaneously outplaying the backbone based networks in security.
To overcome the distance limitations, quantum repeaters (QR) have been proposed in 1998 as a device that allows to distribute entangled particles over arbitrarily long distances~\cite{Briegel1998}.
%The idea is as follows.
The desired distance is divided into shorter intervals as in Fig.~\ref{QRep-fig}. Within every interval entanglement is shared in a standard way by creating entangled particles $A-A'$ and $B-B'$ and distributing them to the interval ends, where photons $A'$ and $B$ are measured jointly, e.g., by a Bell measurement, such that the remaining pair of particles $A$ and $B'$ becomes entangled. This procedure is known as \textit{entanglement swapping}. Since the timing is a crucial factor in this process, the photons have to be stored in a \textit{quantum memory}, where the quantum objects can be stored for a long time without distortion of their states. Various architectures of quantum repeaters have been suggested and various elementary parts of it have been implemented~\cite{Albrecht2014,Hofmann2012,Rielander2014,Saglamyurek2011} and analyzed \cite{Abruzzo2012,Muralidharan2016}.
However, an integrated setup has not yet been finished.
%\todo[inline,color=white]{Das sollte man genauer erklaeren. Was wurde schon gemacht? Was muss noch gemacht werden?}

%!TEX root = main.tex

\section{Challenges \& Outlook}
\label{sec.outlook}

We summarized the current state of realizing long-term secure communication on the Internet on the basis of quantum technology.
To achieve this goal, practical quantum-based key distribution protocols that can also be used in a multi-user network setting are required.
We have discussed the functionality, performance, and security of the most prominent candidates for quantum-based key distribution protocols.
They currently allow for data rates of approximately \SI{20}{\kilo\bit\per\second} over a distance of \SI{100}{\kilo\meter}.
A maximum communication distance of \SI{404}{\kilo\meter} can be achieved at a data rate of \SI{e-4}{\bit\per\second}.
Quantum-based key distribution in multi-user networks is currently realized using relay nodes, which must be trusted to not compromise the confidentiality of the transmitted data.
Furthermore, we discuss an approach for realizing quantum hubs that allow for quantum-based key distribution in multi-user networks without trusted nodes.
We also briefly summarize work on quantum repeaters that would allow to remove the distance limitations of current QKD technology.

In view of the current state of QKD technology, we identify the following open challenges for realizing QKD-based long-term secure communication on the Internet:
\begin{itemize}
\item The data rate of QKD protocols needs to be further improved so that comparable data rates as in classical communication can be achieved.
\item Candidate QKD protocols need to be identified that allow for a secure implementation resistant to known theoretical and practical attacks.
\item Secure connection protocols (e.g., TLS) need to be re-designed to support QKD-based information-theoretically secure key distribution.
Understanding is furthermore needed in how to combine information-theoretic confidentiality with integrity and message-ordering protection in a channel protocol, taking into account real-world aspects like message fragmentation or bi-directional communication.
\item The proposed approaches for realizing quantum hubs need to be implemented and their practicality has to be shown.
\item The practicality of quantum repeaters needs to be shown in implementations and it must be shown how they can be combined with quantum hubs in order to allow for QKD in a wide area multi-user network like the Internet.
\end{itemize}

Progress towards a wider deployment of QKD was made this year, in the shape
of initial experiments performing QKD communication or preparing it with satellites~\cite{Guenthner2017,Takenaka2017}. These initiatives suggest the possibility of a satellite-based free-space network, where satellites are trusted nodes and out of reach for attackers. Alternatively, drones could be used to create such a network~\cite{Sasaki2017}.
For the future, it is envisioned that quantum channels are combined to realize the so called quantum Internet~\cite{Kimble2008}.
Such an Internet infrastructure would constitute a significant advancement in quantum information processing allowing for novel applications, such as quantum secret sharing~\cite{Nascimento2001} and distributed quantum computation, which so far have only been explored theoretically.

\iffalse
\todo[inline]{Networks using drones, analogously to Internet distributing drones by Facebook}
\todo[inline]{O,A: Should we point out that different layers of Internet require different performance of the systems? Nodes of wide area networks should possess higher bandwidth as nodes in metropolitan area.}

\clearpage
\todo[inline]{ask Felix about QKD and OTP Encryption for TLS}

\todo[inline]{Weitere Themen, die angesprochen werden sollen: \\ -mehr zu quantum secret sharing, was wird damit möglich sein. Multi party computation?\\
-quantum computer\\
-Quantum secret sharing\\
-quantum Internet, purpose and applications. It won't be an alternative to classical Internet\\
-quantum secure authentication\\
-high dimensional QKD~\cite{Bacco2016,Mower2012,Niu2016}\\
-Quantum coin flipping\\
-Quantum commitment as a step towards oblivious transfer and secure multi party computation\\
-Cite Lincos\\
-write 'information-theoretic' or 'information theoretic'?}

{Quantum secret sharing}: A generalization of QKD is a quantum secret sharing, that would enable secret distribution of quantum information. In that case
information about a quantum state or a secret key can be shared between several parties. Analogously to classical secret sharing, each party alone cannot restore the whole secret. Already a set-up for E91 protocol can be used for quantum secret sharing \cite{}.
\fi

\section*{Acknowledgment}

This work has been co-funded by the DFG as part of projects P4, S4, and S6 within the CRC 1119 CROSSING.

\bibliographystyle{splncs03}
\bibliography{bibl}

\end{document}